\documentclass[reprint, amsmath,amssymb, aps,nofootinbib]
{revtex4-1} 

\usepackage{graphicx}
\usepackage{dcolumn}
\usepackage{bm}
\usepackage{mathrsfs}
\usepackage{dsfont}
\usepackage{braket}
\usepackage[hidelinks]{hyperref} 
\usepackage{xcolor}
\hypersetup{colorlinks,
	linkcolor={red},
	citecolor={blue},
	urlcolor={green}}
   
\newcommand{\im}{\mathrm{i}}
   
\begin{document}
\title{Superfluidity from correlations in driven boson systems}

\author{Jes{\'u}s Mateos}
\author{Charles Creffield}
\author{Fernando Sols}
 \email{f.sols@ucm.es}
 \affiliation{Departamento de F\'isica de Materiales, Universidad
Complutense de Madrid, E-28040 Madrid, Spain}

\begin{abstract}
We investigate theoretically the superfluidity of a one-dimensional boson system whose hopping energy is periodically modulated with a zero time average, which results in the suppression of first-order single-particle hopping processes. The dynamics of this Floquet-engineered flat-band system is entirely driven by correlations and described by exotic Hamiltonian and current operators. We employ exact diagonalization and compare our results with those of the conventional, undriven Bose-Hubbard system. We focus on the two main manifestations of superfluidity, the Hess-Fairbank effect and the metastability of supercurrents, with explicit inclusion of an impurity when relevant. Among the novel superfluid features, we highlight the presence of a cat-like ground state, with branches that have opposite crystal momentum but carry the same flux-dependent current, and the essential role of the interference between the collective components of the ground-state wave function. Calculation of the dynamic form factor reveals the presence of an acoustic mode that guarantees superfluidity in the thermodynamic limit.
\end{abstract}

\maketitle

\section{\label{sec:intro}Introduction}

Most transport phenomena rely on the presence of particles with the capacity to move individually, as encoded in the kinetic energy term of the microscopic Hamiltonian, or the hopping term in case of lattices. Recently there has been interest in quantum many-body systems whose dynamics is not based on the kinetic energy of individual particles but rather on correlated behavior stemming from interactions. 
For instance, considerable attention has been paid to the Sachdev-Ye-Kitaev (SYK) model for fermions, which postulates two-body interactions that are random and long-ranged in a real-space lattice. This not only provides a route to the study of metallic behavior without quasiparticles \cite{22CH}, but has also been proposed as providing a means of modeling phenomena in quantum gravity \cite{sycamore}.

Both in solid state and cold atom physics, flat band systems have attracted considerable attention in recent years. In the case of quantum materials, this research has been spurred by the discovery of superconductivity in twisted bilayer graphene \cite{18CA} at the magic angles where a flat band is predicted to occur \cite{11BI}.

Systems with a vanishing single-particle group velocity have also been investigated in the context of cold atom setups and for bosons in particular \cite{10HU,20ZU}. The origin of the single-particle flat band can be due to frustration \cite{10HU}, spin-orbit coupling \cite{17HU}  or, more often, to the destructive interference between different paths intervening in the elementary hopping process, giving rise to what is termed ``Aharonov-Bohm caging'' \cite{20ZU,13HY,13TA,18TO,20KU,20PE}.

Recently, the use of Floquet engineering has been proposed to realize many-particle systems in which first-order (i.e., unassisted single-particle) hopping processes are suppressed \cite{18PI, 19PI, 21MA}, which provides an alternative path to the design of flat-band systems. The method relies on a fast time-periodic modulation of the hopping energy with a vanishing time average. 
When applied to the Bose-Hubbard model
we have shown that this kinetic driving leads to an atypical interacting many-body system without a single-particle hopping term \cite{18PI}.
As a function of the driving amplitude, the system can be tuned continuously from the Mott-insulating regime to a peculiar form of superfluidity where the system
shows a relatively robust cat-like ground state with branches characterized by the preferred occupation of opposite nonzero momentum eigenstates \cite{19PI}. In the absence of a flux threading the ring, or in a segment delimited by hard walls, the two cat branches carry no current (here described by another exotic operator \cite{19PI}) but their different crystal momentum makes them behave very differently when allowed to expand in a larger lattice \cite{22MO}. The main properties of this system have been shown to be robust against variations in the driving signal and switching protocol \cite{21MA}.

The purpose of the present work is to explore the superfluidity of the kinetically driven Bose-Hubbard (KDBH) model, of which so far we have presented indirect evidence based on general theorems for one-dimensional systems \cite{18PI, 12SO, 11CA, 03GI} and hints of its unusual nature \cite{19PI}. As a benchmark, and using the same exact diagonalization method, we study in parallel the superfluidity of the conventional (i.e., undriven) Bose-Hubbard (CBH) model, a relatively well-understood system which however still can offer some new insights, especially when treated exactly.

Superfluidity and superconductivity have long been recognized as fundamental quantum phenomena of an essentially identical nature, both involving some form of Bose-Einstein condensation and only differing in the charge of the elementary carriers, which can be bosons or paired fermions. As emphasized by Leggett \cite{06LE}, the term superconductivity actually refers to a variety of phenomena that usually come together and which roughly amount to the Meissner effect (an equilibrium property) and the metastability of supercurrents (a non-equilibrium property). A similar classification applies to neutral superfluids, albeit with nuances that are discussed in section \ref{sec:definitions}. In this paper we investigate both aspects of superfluidity in the KDBH model, whenever possible comparing with the CBH system \footnote{We note that in one dimension the Mermin-Wagner theorem formally
forbids the existence of Bose condensation. For convenience, however,
when discussing these one dimensional systems we shall generally use the
term ``condensate'' in preference to the more correct but verbose ``quasi-condensate''.}.
Interestingly, a system of independent bosons is superfluid only in one of these senses and, for this reason, in our discussion we also consider the CBH model in the particular case of zero interaction. 

Common to all phenomena that fit under the umbrella of superconductivity and superfluidity is the resilience of quantum flow against the presence of barriers or impurities of moderate strength. Thus, a major component of our present work is devoted to the generalization of the KDBH model to the presence of an impurity in the one-dimensional ring.

By explicitly considering the presence of an impurity and the full current-flux dependence, as well as the excitation spectrum (see below), our theoretical proof of superfluidity goes beyond the calculation of the condensate fraction or the Drude weight invoked in e.g. Refs. \cite{17HU,18TO} to claim superfluidity in flat-band boson systems.

This paper is arranged as follows. Section \ref{sec:transport} discusses the central concept of quantum transport from correlations. In section \ref{sec:definitions} we describe the two main notions of superfluidity as addressed in the present context. Section \ref{sec:impurity-model} presents the model Hamiltonian and current operator used in this work. In section \ref{sec:HF} we present and discuss the numerical results for the Hess-Fairbank effect, which amounts to an incomplete Meissner effect due to the absence of Amp\`ere's law for neutral particles. Section \ref{sec:persistent} deals with a discussion of metastable supercurrents and its possible decay mechanisms. It includes a calculation of the dynamic form factor. In both sections \ref{sec:HF} and \ref{sec:persistent} we argue that the superfluidity of the kinetically driven Bose-Hubbard system is, in many regards, qualitatively different from that of the conventional BH model. Section \ref{sec:conclusions} contains some concluding remarks.

\section{\label{sec:transport}Quantum transport from correlations}

In a tight-binding lattice with periodic boundary conditions, the site and momentum boson operators are related by
\begin{equation}
	a_{k_\ell} = \frac{1}{\sqrt{L}}\sum_{x=0}^{L-1} e^{\im k_\ell x}a_{x}\,,\quad a_{x} = \frac{1}{\sqrt{L}}\sum_{\ell=0}^{L-1} e^{-\im k_\ell x}a_{k_\ell}\, ,
	\label{eq:pw_expansion}
\end{equation}
where the crystal momenta take values $k_\ell=2\pi\ell/L$, with $\ell\in \{0,..., L-1\} \subset \mathds{Z}$.

In systems where the total particle number is conserved, the fundamental requirement for the possibility of quantum transport is $\dot \rho_q \neq 0$, where $\dot\rho_q = [\rho_q, H]/i\hbar$ is the time-derivative of the Fourier component of the density operator, 
\begin{equation}
	\rho_q = \sum_x n_x e^{iqx}=
	\sum_k a_k^\dagger a_{k+q} \, ,
	\label{rho-q}
\end{equation}
with $n_x=a_x^\dagger a_x$, and $H$ the underlying Hamiltonian.

In a typical many-body problem with two-body density-density interactions of the type
\begin{equation}
	H_{\rm int}=\frac{1}{2 \Omega}\sum_q \rho_q V_q \rho_{-q} \, .
	\label{H-int}
\end{equation}
where $V_q$ is the Fourier transform of the two-body interaction potential $V(x-x')$, and $\Omega \rightarrow \infty$ is the thermodynamic volume, we have
\begin{equation}
	\dot{\rho}_q = [\rho_q, H_{\rm int}]/i \hbar = 0
	\label{rho-dot}
\end{equation}
Therefore, a kinetic energy term (or hopping energy in a tight-binding picture) is needed to have $\dot \rho_q \neq 0$ and, with it, the possibility of quantum transport.

The Hamiltonians \eqref{eq:Hkdbh}-\eqref{eq:Hkdbh2} and \eqref{H-Uwxyz} which we will derive in section \ref{sec:impurity-model} have in common that they do not commute with the density operator and thus give rise to a highly correlated form of quantum transport, qualitatively different from the conventional transport supported by a standard kinetic-energy term or, in a tight-binding picture, by the elementary process of unassisted single-particle hopping between neighboring sites.

For neutral bosons, the Hamiltonian \eqref{eq:Hkdbh} gives rise to a novel type of superfluidity, microscopically based on the correlated motion of particles and whose study is the object of this article.

\section{\label{sec:definitions}Definitions of superfluidity}

Superfluidity is generally associated with the capacity of a quantum fluid to flow through barriers without a drop in the chemical potential. However, when we need to be more specific, and as noted in the Introduction, we must recognize that superfluidity actually comprises a variety of phenomena that usually (but not universally) go together. In the case of superconductors, these phenomena essentially boil down to two: the Meissner effect, an equilibrium effect under the constraint of a magnetic field, and the existence of persistent currents, a non-equilibrium property \cite{06LE}.

For superfluids, the equivalent of a magnetic field is a rotation and the absence of electric charge translates into the Hess-Fairbank (HF) effect, which may be viewed as an incomplete Meissner effect where only the London equation applies and the equivalent of Amp\`ere's law is missing because its gravitational equivalent, the Lense-Thirring effect, is extremely weak \footnote{We might identify a third phenomenon which is the flow of current through a capillary connecting two bulk reservoirs. As argued in Ref. \cite{06LE}, the physics of such a setup amounts to the HF effect or the metastability of supercurrents depending on whether the total phase variation along the constriction is smaller or larger than $\pi$, respectively.}. 

In our model system, the HF effect is studied by calculating the dependence of the space- and time-averaged current on the external flux $L\phi$ and in particular its linear dependence for small flux. This is the object of Section \ref{sec:HF}, where the HF effect is studied in the presence and absence of an impurity for both interacting and independent bosons.

The existence of metastable supercurrents is explored in Section \ref{sec:persistent} by investigating the possible decay mechanisms for a flow without pressure drop through a barrier or impurity without the assistance of an external flux (whose presence is essential for the HF effect). To that end we focus on the ``mean-field current-carrying excited states'', which are those configurations characterized by the macroscopic occupation of a state with nonzero momentum and nonzero current.

For a single barrier in the ring, the lifetime against current reversal diverges as $\sim L$ for a single particle in the thermodynamic limit, and even more strongly for $N$ repulsively interacting bosons, with $N$ large.

The other decay mechanism is the spontaneous generation of quasiparticles when the speed of flow exceeds the Landau critical velocity. The existence of an acoustic mode guarantees superfluidity for flow speeds below the speed of sound. The Landau decay mechanism, when it applies, becomes most relevant in the large $L$ limit, because a continuum of possible momenta permits the excitation of quasiparticles of arbitrarily low momentum and energy \footnote{In quasi--one-dimensional systems, phase slips can be another current decay mechanism. We do not address it here.}.

In Ref. \cite{18PI}, the superfluidity of the KDBH system was indirectly proved by invoking general theorems from Luttinger liquid theory \cite{18PI, 12SO, 11CA, 03GI}. Here we will perform a more direct check by computing the dynamic form factor \cite{96PI}, showing that the results obtained are consistent with the presence of an acoustic mode and thus a nonzero Landau critical velocity.

Both for the HF effect and the metastability of supercurrents, we include calculations for the undriven, conventional BH system to use it as a benchmark.

\section{\label{sec:impurity-model}The impurity model}

\begin{figure*}
\centering
\includegraphics[width=0.7\linewidth]{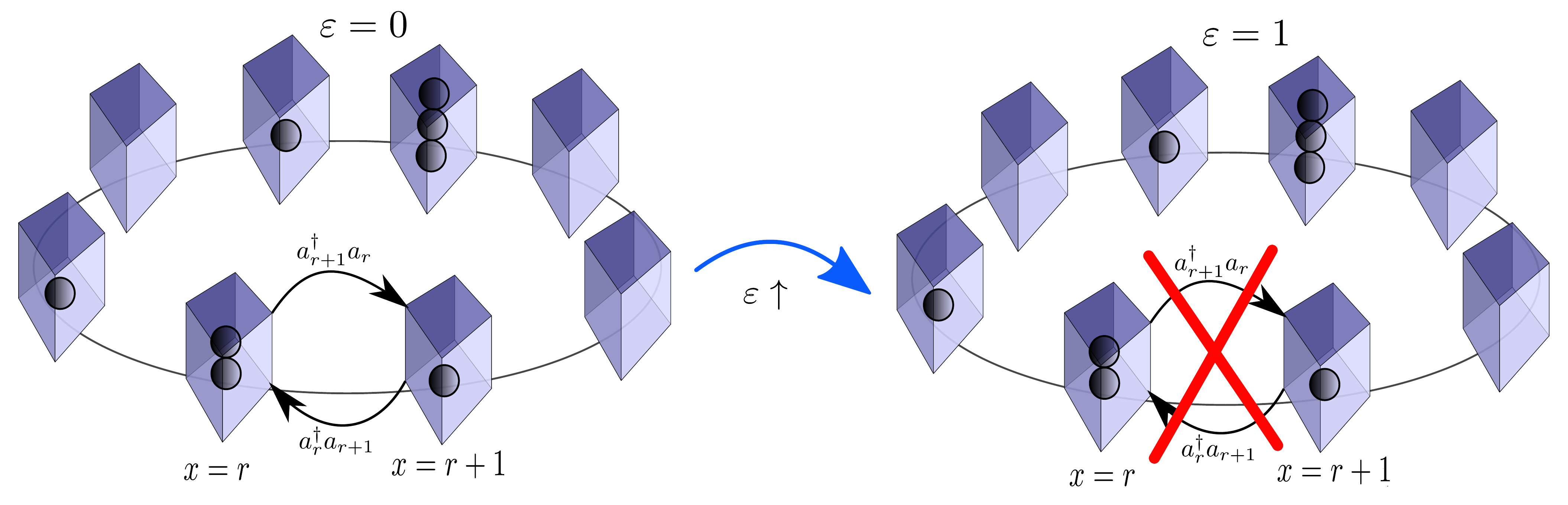}
\caption{Off-diagonal impurity in a ring located at the link between sites $r$ and $r+1$. As the impurity strength $\varepsilon$ increases from 0 to 1, the link between the two adjacent sites is broken and we recover the hard-wall limit.}
\label{fig:impurity} 
\end{figure*}

Since we are interested in superfluidity, we generalize the work of Refs. 
\cite{18PI, 19PI,21MA} to include an impurity and a threading flux in the BH ring, a setup which in the absence of driving can be modeled with the Hamiltonian
\begin{align}
	\nonumber
	\mathcal{H}=&-J\sum_{x=0}^{L-1}\left(1-\delta_{rx}\varepsilon\right)\left(e^{i\phi}a^\dagger_{x+1}a_x+\text{H.c.}\right)\\\label{eq:Hcbhpos}
	&+\frac{U}{2}\sum_{x=0}^{L-1}n_x(n_x-1) \, .
\end{align}
This is the periodic 1D Bose-Hubbard Hamiltonian (we impose $a_x=a_{x+L}$) with an external flux per link $\phi$ and an off-diagonal impurity parameterized by $\varepsilon\in[0,1]$. We assume $U > 0$.
For $\varepsilon=0$ the system is a periodic ring. As $\varepsilon$ increases, the hopping amplitude between sites $x=r$ and $x=r+1$ reduces, as schematically depicted in Fig. \ref{fig:impurity}. The limit $\varepsilon = 1$ corresponds to the segment or hard-wall limit, which has been investigated in Ref. \cite{19PI}. In practice a continuous value of $\varepsilon$ can be realized by shining a blue-tuned laser into a small section of the ring in order to erect a repulsive barrier of tunable height.

From a numerical viewpoint, a practical advantage of the off-diagonal feature is that the
system can be continuously tuned from a ring to a segment by varying $\varepsilon$ 
without changing the number of sites. Hereafter we set $\hbar = 1$ and momenta will be measured in units of the inverse lattice spacing.

In the momentum representation, \eqref{eq:Hcbhpos} acquires the form
\begin{align}
	\nonumber
	\mathcal{H}=&-2J\sum_{\ell=0}^{L-1}\cos(k_\ell+\phi)a^\dagger_{k_\ell}a_{k_\ell}
	\\\nonumber
	&+J\frac{\varepsilon}{L}\sum_{\ell, m=0}^{L-1}h(k_\ell, k_m; \phi)a^\dagger_{k_\ell}a_{k_m}\\ \label{eq:Hcbh}
	&+\frac{U}{2L}\sum_{\ell, m, n, p=0}^{L-1}\delta_{k_\ell+k_m, k_n+k_p}a^\dagger_{k_p} a^\dagger_{k_n}a_{k_m}a_{k_\ell} \,
\end{align}
where the function $h$ reads
\begin{equation}
	\nonumber
	h(k_\ell, k_m; \phi)=e^{i\left(k_\ell-k_m\right)\left(r+\frac{1}{2}\right)}\cos\left(\frac{k_\ell+k_m}{2}+\phi\right) \, .
	\label{eq:h}
\end{equation}
The presence of a Peierls phase $\phi$ permits the inclusion of an effective total flux $L\phi$ threading the ring.


Under the effect of fast kinetic driving, $J\to J\cos(\omega t)$, and as described in Refs. \cite{18PI,19PI}, an effective static Hamiltonian 
$H_{\rm eff}$ results which can be written in closed form only for small $\varepsilon$, as indicated below, 
where the result is given up to order $\varepsilon^2$. The hard-wall limit $\varepsilon=1$ also admits a closed form, as shown in Ref. \cite{19PI}. 

The smallness of $\varepsilon$ is not a major shortcoming because, to investigate superfluidity, it suffices to study the case of weak impurities. Specifically, we calculate
\begin{equation}
\label{eq:Hkdbh}
H_{\text{eff}}=H^{(0)}_{\text{eff}}+H^{(1)}_{\text{eff}}+H^{(2)}_{\text{eff}}+\mathcal{O}(\varepsilon^3).
\end{equation}
	
The zeroth-order term $H^{(0)}_{\text{eff}}$ (absence of impurity) reads
\begin{align}
	\nonumber
	H^{(0)}_{\text{eff}} = & \frac{U}{2L}  \sum_{\ell, m, n, p=0}^{L-1}\delta_{k_\ell+k_m, k_n+k_p} \\
	& \times \mathcal{J}_{0}\left[2\kappa  
	F(k_\ell,k_m,k_n,k_p;\phi)\right] a^\dagger_{k_p}
	a^\dagger_{k_n}a_{k_m}a_{k_\ell}\, ,
	\label{Hkdbh0}
\end{align}
where $\mathcal{J}_0$ is the zeroth order Bessel function of the first kind, $\kappa=J/\omega$ is the driving parameter, and
\begin{align}
	\nonumber
	F(k_\ell,k_m,k_n,k_p;\phi)=\phantom{-}&\cos(k_\ell+\phi)+\cos(k_m+\phi)\\
	\label{eq:Ffunction}
	-&\cos(k_n+\phi)-\cos(k_p+\phi).
\end{align}
The resulting impurity-free ring model has been studied in Refs. \cite{18PI,19PI} for $\phi=0$ and in Ref. \cite{21MA} for nonzero $\phi$. Having been obtained by Floquet analysis in the high frequency limit \cite{18PI} it can be understood as the first term in an expansion in inverse powers of $\omega$, such as the Magnus or van Vleck series. In principle the system's behaviour could be described at lower driving frequencies by calculating higher terms in the series expansion, or by applying alternative forms of Floquet engineering valid over all frequencies, such as that described in Ref. \cite{lie_algebra}. In this work, however, we simply confine ourselves to the $\omega \rightarrow \infty$ limit.

Within the same order in the inverse-frequency expansion, the terms of higher-order in the impurity strength are:
\begin{widetext}
	\begin{align}
		\label{eq:Hkdbh1}
		H^{(1)}_{\text{eff}}&=-\frac{U\kappa^2\varepsilon}{2L^2}\sum_{\ell, m, n, p, v=0}^{L-1}\delta_{k_\ell+k_m, k_n+k_p}[F(k_\ell,k_m,k_n,k_p;\phi)+F(k_\ell,k_m, k_v, k_p;\phi)]h(k_v, k_n;\phi)a^\dagger_{k_p}a^\dagger_{k_v}a_{k_m}a_{k_\ell}+\text{H.c.,}\\
		\nonumber
		H^{(2)}_{\text{eff}}&=-\frac{U\kappa^2\varepsilon^2}{4L^3}\sum_{\ell, m, n, p, v, w=0}^{L-1}\delta_{k_\ell+k_m, k_n+k_p}h(k_v,k_n;\phi)\left[h(k_w, k_v;\phi)a^\dagger_{k_p}a^\dagger_{k_w}a_{k_m}a_{k_\ell}
		+h(k_w, k_p;\phi)a^\dagger_{k_v}a^\dagger_{k_w}a_{k_m}a_{k_\ell}\right.\\
		\label{eq:Hkdbh2}
		&\phantom{aaaaaaaaaaaaaaaaaaaaaaaaaaaaaaaaaaaaaaaaa}\left.-2h(k_\ell, k_w;\phi)a^\dagger_{k_p}a^\dagger_{k_v}a_{k_m}a_{k_w}\right]+\text{H.c.}\, 
	\end{align}
\end{widetext}

\subsection{\label{subsec:current}Current operator}

The space-averaged particle current operator
\begin{equation}
\label{I-H-phi-general}
\mathcal{I}=\frac{1}{L}\frac{\partial \mathcal{H}}{\partial \phi}
\end{equation}
for the CBH case is given from \eqref{eq:Hcbh} by
\begin{align}
	\nonumber
	\mathcal{I}=&\frac{2J}{L}\sum_{\ell=0}^{L-1}\sin(k_\ell+\phi)a^\dagger_{k_\ell}a_{k_\ell}\\
	\label{eq:Icbh}
	&+J\frac{\varepsilon}{L^2}\sum_{\ell, m=0}^{L-1}\tilde h(k_\ell, k_m; \phi)a^\dagger_{k_\ell}a_{k_m},
\end{align}
where now
\begin{align}
	\label{eq:htilde}
	\tilde h(k_\ell, k_m; \phi)=e^{i\left(k_\ell-k_m\right)\left(r+\frac{1}{2}\right)}\sin\left(\frac{k_\ell+k_m}{2}+\phi\right).
\end{align}

Starting from \eqref{eq:Icbh}, and with manipulations similar to those leading from \eqref{eq:Hcbh} to \eqref{eq:Hkdbh} \cite{18PI} (but somewhat more involved because of the flux dependence of the intermediate canonical transformation), we obtain the following time-averaged current operator in the KDBH case:
\begin{widetext}
	\begin{align}
		\label{eq:Ikdbh}
		I_{\text{eff}}
		&
		= I^{(0)}_{\text{eff}}+I^{(1)}_{\text{eff}}+I^{(2)}_{\text{eff}}+\mathcal{O}(\varepsilon^3),\\
		\label{eq:Ikdbh0}
		I^{(0)}_{\text{eff}}&=\frac{U\kappa}{L^2}\sum_{\ell, m, n, p=0}^{L-1}\delta_{k_\ell+k_m, k_n+k_p}\mathcal{J}_1\left[2\kappa F(k_\ell,k_m,k_n,k_p;\phi)\right]G(k_\ell,k_m,k_n,k_p;\phi)a^\dagger_{k_p}a^\dagger_{k_n}a_{k_m}a_{k_\ell},\\
		\nonumber
		I^{(1)}_{\text{eff}}&=\frac{U\kappa^2\varepsilon}{2L^3}\sum_{\ell, m, n, p, v=0}^{L-1}\delta_{k_\ell+k_m, k_n+k_p}\Big\{\left[G(k_\ell,k_m, k_n, k_p;\phi)+G(k_\ell,k_m, k_v, k_p;\phi)\right]h(k_v, k_n;\phi)\\\label{eq:Ikdbh1}
		&\phantom{aaaaaaaaaaaaaaaaaaaaaaaa}\;\;+[F(k_\ell,k_m, k_n, k_p;\phi)+F(k_\ell,k_m, k_v, k_p;\phi)]\tilde{h}(k_v, k_n;\phi)\Big\}a^\dagger_{k_p}a^\dagger_{k_v}a_{k_m}a_{k_\ell}+\text{H.c.,}\\\nonumber
		I^{(2)}_{\text{eff}}&=\frac{U\kappa^2\varepsilon^2}{4L^4}\sum_{\ell, m, n, p, v, w=0}^{L-1}\delta_{k_\ell+k_m, k_n+k_p}
		\Big\{\left[\tilde{h}(k_v,k_n;\phi)h(k_w, k_v;\phi)+h(k_v,k_n;\phi)\tilde{h}(k_w, k_v;\phi)\right]a^\dagger_{k_p}a^\dagger_{k_w}a_{k_m}a_{k_\ell}\\\nonumber
		&\phantom{aaaaaaaaaaaaaaaaaaaaaaaaa}\;\,+\;\;\,\,\left[\tilde{h}(k_v,k_n;\phi)h(k_w, k_p;\phi)+h(k_v,k_n;\phi)\tilde{h}(k_w, k_p;\phi)\right]a^\dagger_{k_v}a^\dagger_{k_w}a_{k_m}a_{k_\ell}\\\label{eq:Ikdbh2}
		&\phantom{aaaaaaaaaaaaaaaaaaaaaaaaa}\;\,-\;2\left[\tilde{h}(k_v,k_n;\phi)h(k_w, k_p;\phi)+h(k_v,k_n;\phi)\tilde{h}(k_w, k_p;\phi)\right]a^\dagger_{k_p}a^\dagger_{k_v}a_{k_m}a_{k_w}\Big\}+\text{H.c.,}
	\end{align}
\end{widetext}
\noindent where $\mathcal{J}_1$ is the first order Bessel function of the first kind and
\begin{align}
	\nonumber
	G(k_\ell,k_m,k_n,k_p;\phi)=\phantom{-}&\sin(k_\ell+\phi)+\sin(k_m+\phi)\\\label{eq:Gfunction}
	-&\sin(k_n+\phi)-\sin(k_p+\phi).
\end{align}
\noindent As before, \eqref{eq:Ikdbh0} represents the effective particle current operator in the case without impurity.

It can be shown that \eqref{eq:Ikdbh} can be obtained from \eqref{eq:Hkdbh} through the relation
\begin{equation}
	I_{\text{eff}}=\frac{1}{L}\frac{\partial H_{\text{eff}}}{\partial \phi} \, ,
	\label{current-energy} 
\end{equation}
as expected. 

\subsection{Site representation}

In the site representation, \eqref{Hkdbh0} acquires the form
\begin{equation}
	H^{(0)}_{\text{eff}} =
	\sum_{w,x,y,z=0}^{L-1} U_{wxyz} a_w^\dagger a_x^\dagger a_y a_z
	\label{H-Uwxyz}
\end{equation}
where $U_{wxyz}$ in principle connects all sites, although its amplitude
reduces as the separation of sites increases \cite{18PI}.  
The amplitudes of the matrix elements depend very weakly on the flux while their phases are insensitive to or linearly dependent on the flux as expected in each case (not shown).

The effective Hamiltonian \eqref{H-Uwxyz} formally resembles the SYK model \cite{22CH}, where the matrix elements of the elementary fermion collisions are infinitely long-ranged and random. By contrast, our effective interaction energies reduce with distance, are non-random, and are realizable through a specific prescription \cite{18PI}.

\section{\label{sec:HF}Hess-Fairbank effect}

The HF effect refers to the failure of a superfluid to be dragged by the rotation of the ring where it resides if both are initially at rest. In the rotating frame, the HF is perceived as the establishment of a current due to the presence of a flux threading the ring. Its superconducting equivalent is described in the London equation, in which the electric current is proportional to the vector potential in the London gauge \cite{96TI}. In a neutral superfluid, that translates into a dependence of the ground-state current on the total effective flux ($L\phi$) threading the ring. 

In this section, we compute the current expectation value $\langle I(\phi) \rangle$ in a variety of relevant situations and models. For small flux we find $\langle I(\phi) \rangle\sim \phi$ but, in the general case the current depends nonlinearly on the flux, very much in the way that in a superconductor, even a uniform one, the current density does not always depend linearly on the vector potential \cite{01SA}. 

The space-averaged current operator is given in Eqs. \eqref{eq:Ikdbh} and \eqref{current-energy}. For $\varepsilon=0$ and $\phi=0$ it was shown in Ref. \cite{19PI} that its expectation value is zero in the ground state and for each of the cat branches taken separately. In the following subsections we aim to understand the structure of the full flux dependence of the current expectation value.

\subsection{\label{subsec:matrix-ele} Matrix elements of the current operator}

If we try to compute the expectation value of the current in the ground state, we must deal with matrix elements associated with elementary processes of the type $k_\ell,k_m \rightarrow k_n,k_p$, as indicated in \eqref{eq:Ikdbh0}. It was argued in Ref. \cite{19PI} that the ground state has, to a large extent, a pairing structure, especially for large driving amplitudes $\kappa$. This means that, in the pairing limit and for $\phi=0$, its internal dynamics is dominated by collisions between pairs of total momentum $\pi$ (or, equivalently, $-\pi$). To generalize this result to the case of $\phi\neq 0$, we focus on the matrix elements of the current operator for processes of the type
\begin{equation}
	\left(\frac{\pi}{2}+p+\phi, \frac{\pi}{2}-p+\phi \right) \rightarrow \left(\frac{\pi}{2}+p'+\phi, \frac{\pi}{2}-p'+\phi \right) \, ,
	\label{elementary}
\end{equation}
where the momenta $p,p'$ are constrained by $-\pi \leq p,p' \leq \pi$. 

To fix the language, we group the many configurations contributing to the ground state into three collective components:
(i) {\it ideal cat}, which is $(|N_{\pi/2}\rangle +|N_{-\pi/2}\rangle)/\sqrt{2}$, where $|N_{k}\rangle$ is the state with $N$ particles in momentum $k$;
(ii) {\it shared condensate}, where only momenta $k=\pm \pi/2$ intervene, excluding the ideal cat;
(iii) {\it reduction cloud}, formed by all configurations where at least one occupied momentum is different from $\pi/2$ and $-\pi/2$.
The union of the first two groups is the (fragmented) condensate.

The current matrix element associated to the elementary process \eqref{elementary} is proportional to 
\begin{equation}
{\cal J}_1[(\cos p - \cos p')\sin \phi](\cos p - \cos p')\cos \phi \, .
\label{pairing-contribution}
\end{equation}
This precludes the contribution from processes for which $p=p'$. It is easy to prove that the ideal cat alone cannot contribute to the current, although it can contribute through interference terms with the other two components. 
If one of the intervening pairs in \eqref{elementary} belongs to the fragmented condensate, e.g.  $p=0$, then necessarily we must have $p'\neq 0$ in order to contribute to the current expectation value. This means that the presence of the reduction cloud or the shared condensate is essential to have a nonzero current, even for the ideal ring ($\varepsilon =0$) with an external flux.

As we shift from KDBH to CBH, the ideal cat state is replaced by the single condensate, while the reduction cloud plus the shared condensate become the usual depletion cloud. 
The central role of the reduction cloud and the shared condensate in the generation of a flux dependent current, contrasts sharply with the case of the CBH system, where the condensate does not need to interfere with the depletion cloud, since it can support a nonzero current by itself when assisted by a flux. In particular, a system of independent bosons does exhibit the HF effect.

This remarkable property of the KDBH system which we have just described, together with the very presence of the cat state, reveals that its superfluidity is qualitatively different from that of conventional Bose systems.

Equation \eqref{pairing-contribution} also reveals that the current vanishes for zero flux and that, as a function of $\phi$, it has a periodicity of at least $\Delta \phi=2\pi$, if not smaller. More information can be obtained from the numerical results which we discuss in the next subsection.

\subsection{\label{subsec:HF-numeric} Numerical results}

We use exact diagonalization based on LAPACK routines to treat the many-body interacting system. The existence of a cat-like ground state relies on the presence of momenta $\pm \pi /2$ in the Brillouin zone, which means that $L$ must be a multiple of 4. The first option, $N = L = 4$ is too small to use except as a test case, so the majority of our results are for eight bosons on eight sites ($N = L = 8$),  which is large enough to yield interesting results, and has a Hilbert space of dimension $(N+L-1)!/N! (L-1)! = 6435$. The next size of interest to us would be $N = L = 12$ which has a Hilbert space of dimension 1,352,078. The full diagonalization of a matrix of this size is beyond our computing power, since the presence of the impurity and the external flux remove the symmetries which would allow us to block-diagonalize the Hamiltonian.

\begin{figure*}
	\centering
	\includegraphics[width=1\linewidth]{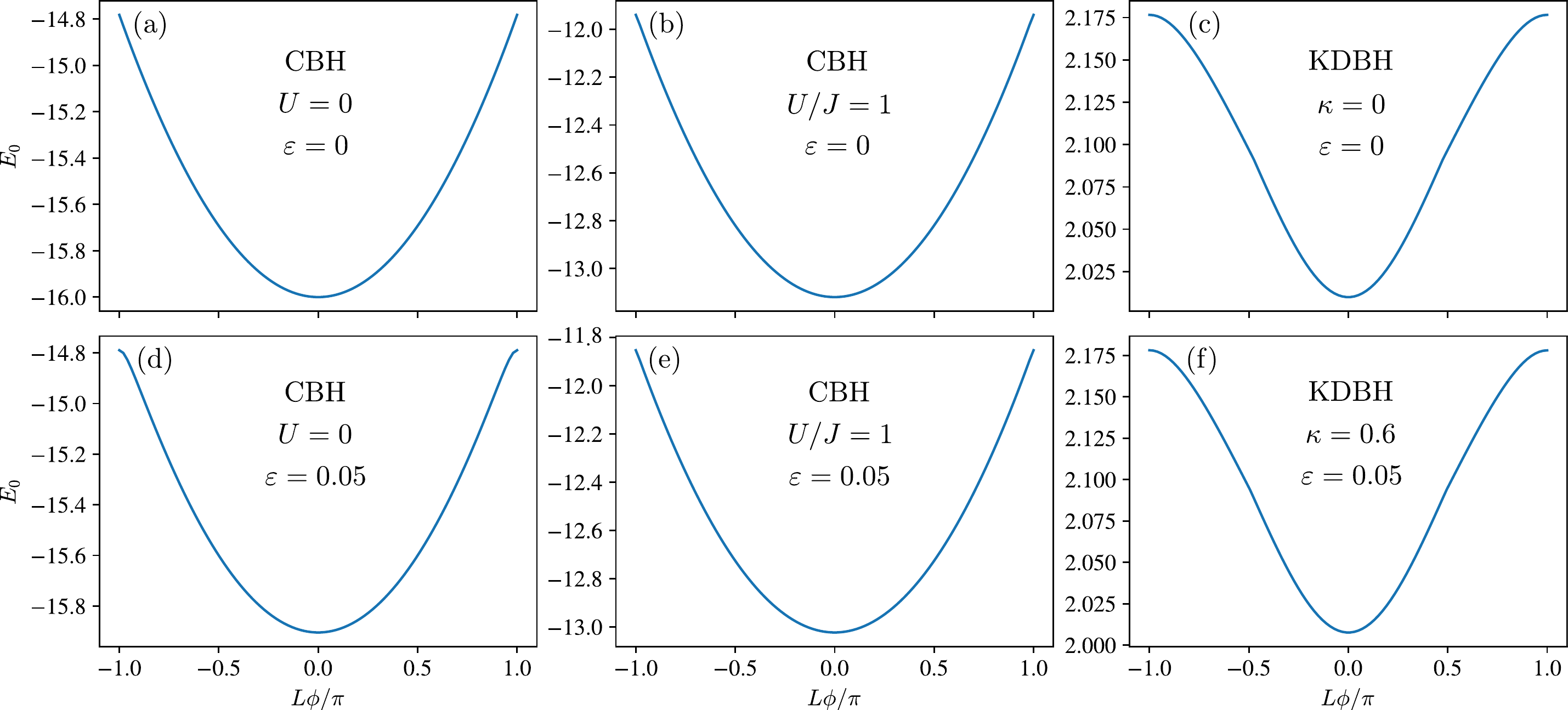}
	\caption{Ground-state energy versus total flux $L\phi$ across the ring for the following cases (always with $J=1$): Schr\"odinger limit [Hamiltonian \eqref{eq:Hcbh} with $U=0$, plots (a),(d)]; CBH [\eqref{eq:Hcbh} in the superfluid regime $U/J=1$], plots (b),(e); KDBH [\eqref{Hkdbh0} in the superfluid regime $\kappa=0.6$], plots (c),(f). The upper row [(a)-(c)] corresponds to an impurity-free system, while the lower row [(d)-(f)] has been calculated with a small but nonzero impurity strength $\varepsilon=0.05$.}
	\label{fig:eigen_vs_phi} 
\end{figure*}

\begin{figure*}
	\centering
	\includegraphics[width=1\linewidth]{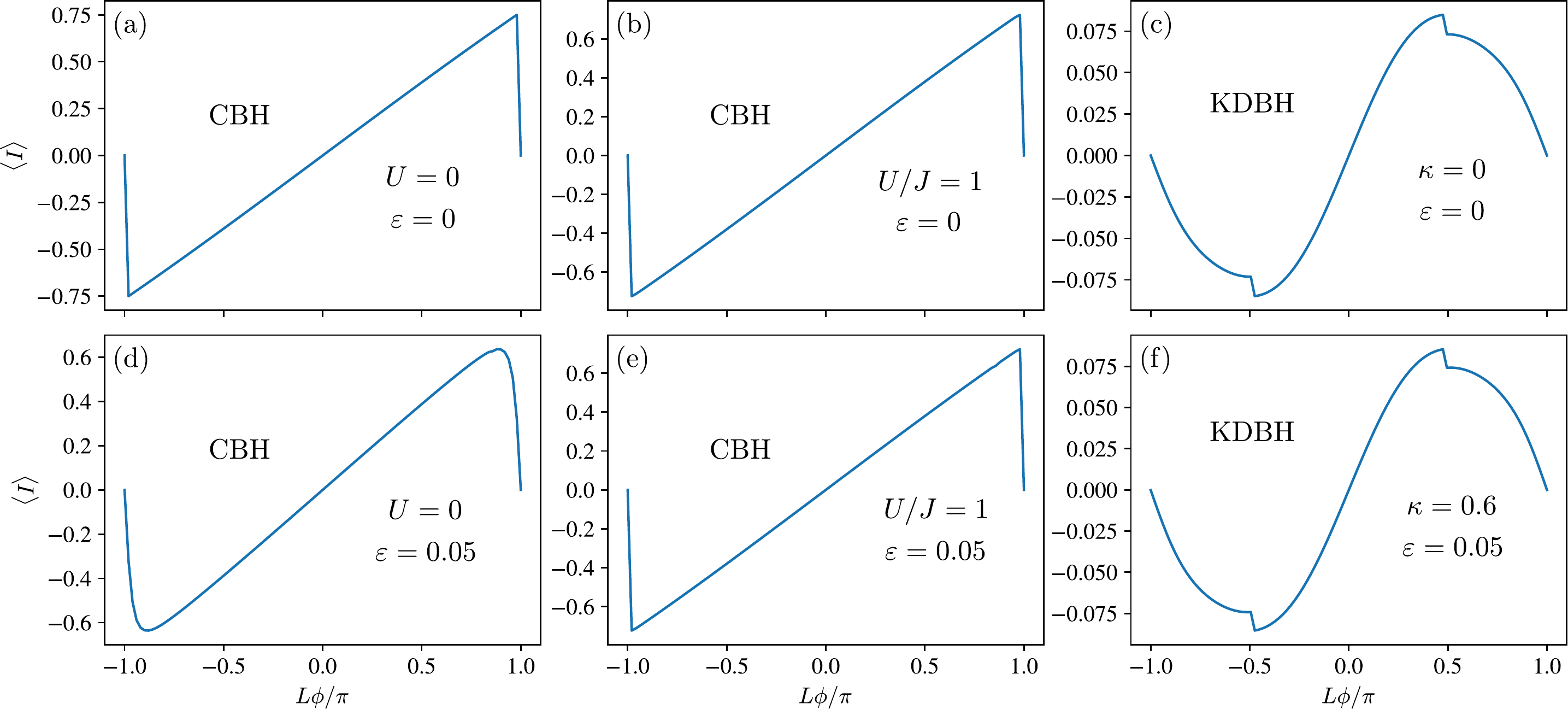}
	\caption{Same as Fig. \ref{fig:eigen_vs_phi}, for the ground-state expectation value of the current operator $\mathcal{I}$ given in \eqref{eq:Icbh} and $I_{\rm{eff}}$ given in \eqref{eq:Ikdbh}
 corresponding to the CBH and KDBH cases, respectively.}
	\label{fig:current_vs_phi} 
\end{figure*}

In Fig. \ref{fig:eigen_vs_phi} we show the energy of the ground state as a function of the flux for the CBH model, both with and without interactions, and for the KDBH system, both for a perfect ring and for a ring with a weak impurity. In all cases we focus on the superfluid regime because the Mott insulating regime is uninteresting for the present purposes. The figures contain some hidden structure that is better appreciated in Fig. \ref{fig:current_vs_phi}, where the the ground state current is plotted as a function of the flux. 

Figures \ref{fig:eigen_vs_phi} and \ref{fig:current_vs_phi} are connected through the general relation \eqref{current-energy}. In Fig. \ref{fig:current_vs_phi} one can see that the properties of the system are periodic in the total external $L\phi$ with periodicity $L\Delta\phi=2 \pi$, as expected from a general theorem due to Bloch for rings of arbitrary dimensionality threaded by a magnetic flux \cite{70BL}. Thus the periodicity of the uniform Peierls phase $\phi$ is $\Delta \phi = 2\pi/L$, a result that cannot be inferred from the discussion in the previous subsection.

For an impurity-free ring, Figs. \ref{fig:current_vs_phi}a,b show that the only effect of interactions is that of slightly decreasing the overall magnitude of the current. On the other hand, Figs. \ref{fig:current_vs_phi}d,e reveal that a weak impurity can change the shape of the curve $I(\phi)$ for independent bosons but not appreciably if there are interactions.

Remarkably, the current as a function of the flux looks qualitatively different in the case of the KDBH system (see Figs. \ref{fig:current_vs_phi}c,f). Like CBH, it is quite insensitive to the presence of a weak impurity. The two discontinuities are due to a crossing between the ground and the first excited state, both with a similar cat structure but differing in the sign with which the two cat branches combine.

\begin{figure*}
	\centering
	\includegraphics[width=1\linewidth]{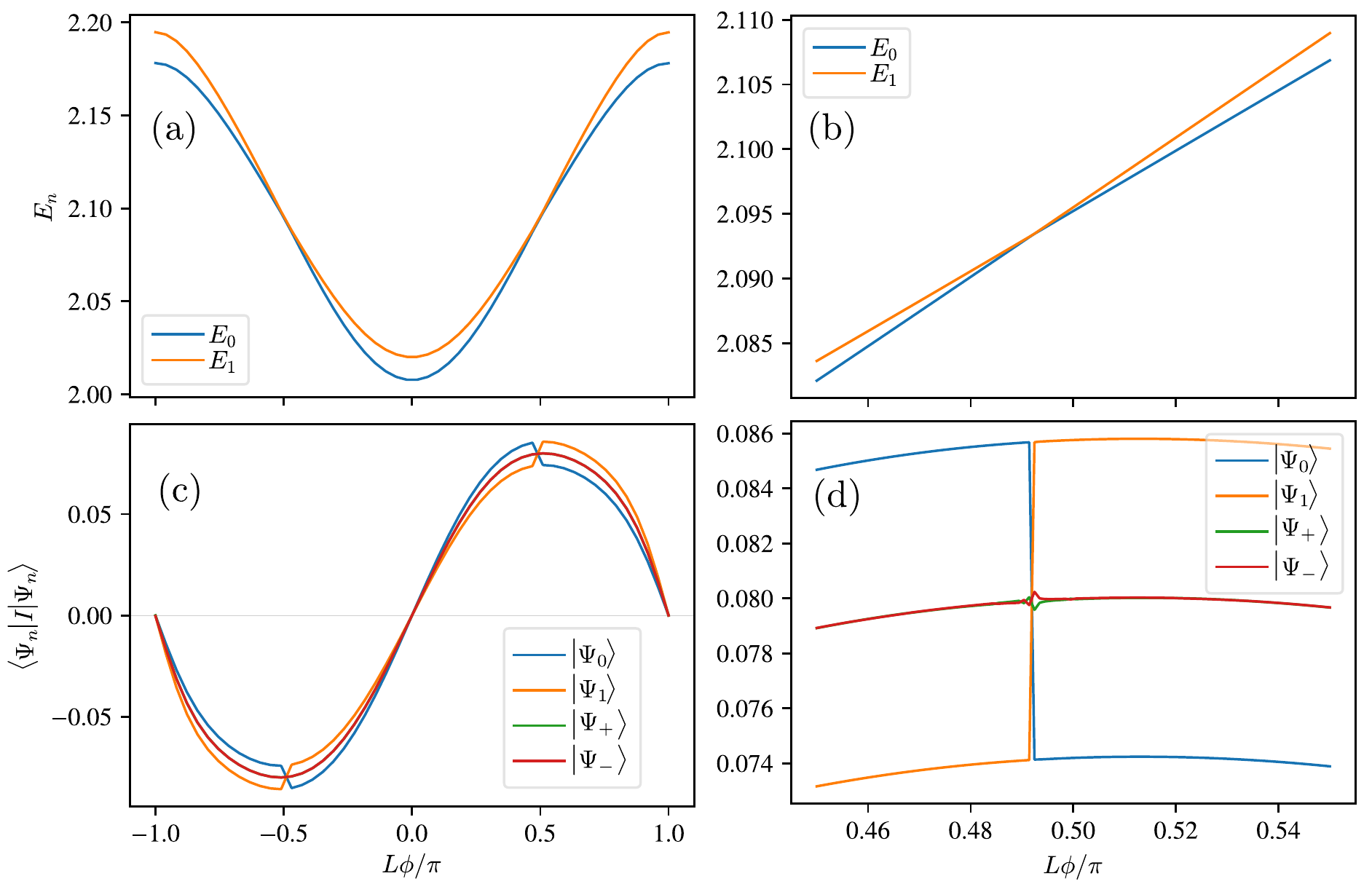}
	\caption{Energy (upper row) and current (lower row) expectation values for the ground state $\ket{\Psi_0}$ and the first excited state $\ket{\Psi_1}$ of the KDBH Hamiltonian \eqref{eq:Hkdbh} as a function of the total flux across the ring $L\phi$. The current expectation values for the separate cat branches $\ket{\Psi_\pm}=(\ket{\Psi_0}\pm\ket{\Psi_1})/\sqrt{2}$ are also plotted and shown to be identical. The right column offers a magnified and finer view of the crossing region in the left column. In all cases, $\kappa = 0.6$, $\varepsilon=0.05$, $N=L=8$.}
	\label{fig:eigen_current_vs_phi_zoom} 
\end{figure*}

This non-avoided crossing can be better appreciated in Fig. \ref{fig:eigen_current_vs_phi_zoom}, where the energy and current of the ground state and the first excited state in the presence of a weak impurity are plotted as a function of the flux in different colors. A zoom of the crossing is shown in Figs. \ref{fig:eigen_current_vs_phi_zoom}b,d.

In the two lower figures (Figs. \ref{fig:eigen_current_vs_phi_zoom}c,d) we also show the energy and current expectation values in the cat branches $|\Psi_{\pm}\rangle$ of the lowest-lying states. Remarkably, despite being characterized by the macroscopic occupation of opposite crystal momenta ($\pm \pi/2$), as discussed in Ref. \cite{19PI},
the two branches carry identical current in the presence of a flux. The two curves actually lie on top of each other in Fig. \ref{fig:eigen_current_vs_phi_zoom}c and differ, but only spuriously, at the magnified crossing of Fig. \ref{fig:eigen_current_vs_phi_zoom}d.

The non-avoided crossing within the ground doublet takes place near but not exactly at $L\phi=\pi/2$. There is no fundamental reason why the crossing should take place at total flux equal to $\pi/2$. Rather, the crossing is due to a delicate dependence of the occupations of momentum states as a function of the flux, resulting in a change of sign in the matrix element connecting the $\pm \pi/2$ cat branches which in this particular case happens to occur near $L\phi=\pi/2$.

\begin{figure*}
	\centering
	\includegraphics[width=1\linewidth]{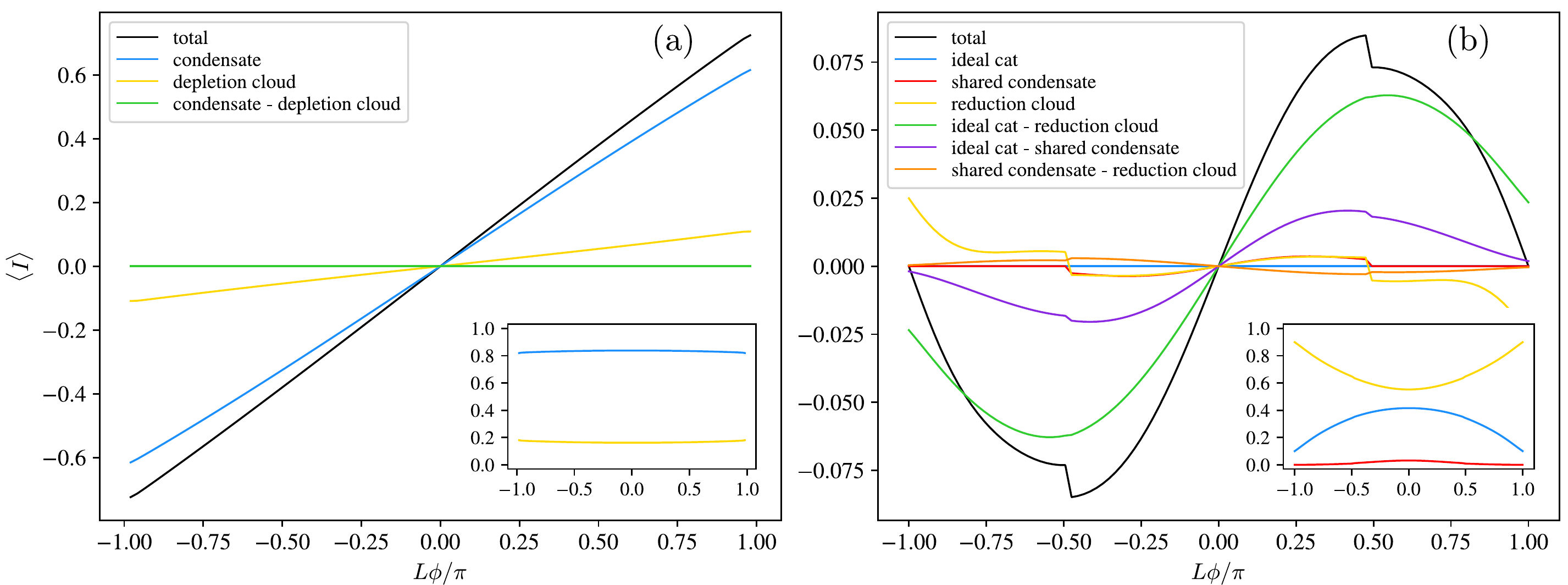}
	\caption{Ground-state expectation value of the current operator for (a) the CBH [Eq. \eqref{eq:Icbh}] and (b) the KDBH [\eqref{eq:Ikdbh0}] systems, as a function of the total flux $L\phi$. The contribution from the different collective components of the ground state is also shown: condensate and reduction cloud for CBH, and ideal cat, shared condensate, and depletion cloud for KDBH, as well as the crossed terms between the components. In both figures, the sum of the colored plots equals the black line. The superfluid regime is characterized by the parameters $J/U=1$ in (a) and  $\kappa = 0.6$ in (b). In all cases $\varepsilon=0$, $N=L=8$. The insets show, with the same color convention, the intrinsic weight of the mentioned ground-state components for the two systems considered.}
	\label{fig:current_condensate} 
\end{figure*}

In Fig. \ref{fig:current_condensate} we show the contribution to the current from the various collective components of the CBH and KDBH ground states including their interference terms. We consider the impurity-free case ($\varepsilon=0$), which is essentially equivalent to the case $\varepsilon=0.05$ shown in Fig. \ref{fig:eigen_current_vs_phi_zoom}. For the KDBH system (Fig. \ref{fig:current_condensate}b), we notice that the current is mostly dominated by the interference of the ideal cat with the shared condensate and the reduction cloud. By contrast, the diagonal contributions from each of the three sectors is quite small. Notably, and as predicted, the diagonal contribution from the ideal cat is zero.

This is in marked contrast with the pattern found for the conventional, undriven system (Fig. \ref {fig:current_condensate}a). There the current is dominated by the diagonal contribution of the condensate, with a small contribution from the depletion cloud and a vanishing interference between the two components.

The insets show the intrinsic weight of the three components within the ground state. In the KDBH case it is interesting to note that the largest probability resides in the reduction cloud but this gives a very small diagonal contribution to the current.

We note that, for $L\phi=\pi$, the momenta are displaced by half their relative spacing. This gives rise to somewhat anomalous behavior for the CBH system in the partial contributions from the condensate and the depletion cloud, although not in the total current, which remains identical to the curve in Fig. \ref{fig:current_vs_phi}b. To avoid confusion, we have removed the points $L\phi/\pi=\pm 1$ in Fig. \ref{fig:current_condensate}a.

\subsection{\label{subsec:n-dependence} Density dependence}

\begin{figure}
	\centering
	\includegraphics[width=1\linewidth]{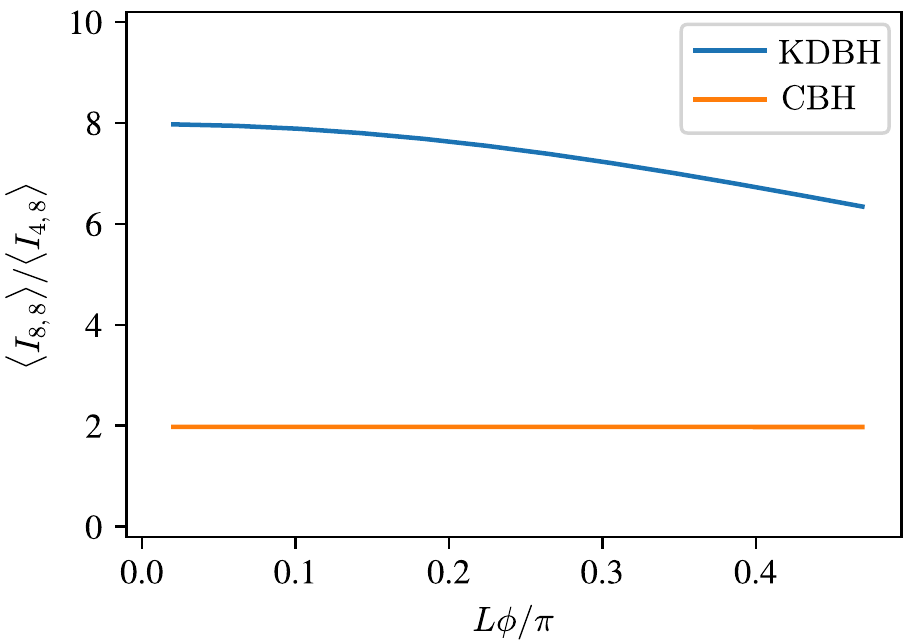}
	\caption{Ratio of the ground-state expectation values of the current for total particle numbers $N=8$ and $N=4$ with a fixed number of sites $L=8$, versus the total flux across the ring $L\phi$, for the CBH and the KDBH systems. The impurity strength is $\varepsilon=0$, while $J/U=1$ and $\kappa=0.6$ for the CBH and KDBH cases, respectively.}
	\label{fig:ratio_current} 
\end{figure}

Another interesting feature of the KDBH system is the unusual density dependence of the current in the presence of a flux. In the conventional BH model, the HF effect is mostly due to the current carried by the condensate, but there is also a contribution from the depletion cloud that further contributes to this effect. Thus one expects the current to scale linearly with the density. This is what in fact we find in Fig. \ref{fig:ratio_current}, where the current magnitude doubles as we go from $(N,L)=(4,8)$ to $(N,L)=(8,8)$. Similar results are also obtained for a weak impurity.

By contrast, the dependence of the current on the density in the KDBH system shows a very different behavior. As the density is doubled, the current increases by a weakly flux-dependent factor which is neither two (as for non-interacting bosons) nor four, as one would naively expect if a mean-field picture were applicable. This non-standard behavior of the current further reflects the fact that the current operator derives from an exotic Hamiltonian describing highly correlated particles.

In this context, it is worth pointing out that, although the current scale is substantially lower in the KDBH system as compared with the CBH model (see Figs. \ref{fig:current_vs_phi} and \ref{fig:current_condensate}) this is not necessarily the case for higher boson densities that are beyond the reach of exact numerical diagonalization.

\subsection{\label{subsec:f_s} Superfluid fraction}

An alternative approach for studying the density dependence of the current relies on the calculation of the superfluid fraction, which in the present context is defined as
\begin{equation}
f_s \equiv \frac{1}{2Jn} \left. \frac{\partial \langle I \rangle}{\partial \phi}\right|_{\phi=0} \, ,
\end{equation}
where $n=N/L$, which is proportional to the Drude weight \cite{03GI}.
For a system of free bosons, $f_s=1$, and the same result should apply for the CBH model in the superfluid regime \cite{98LE}. Numerically we obtain essentially perfect agreement with the value of unity in the non-interacting ($U=0$) case, both for (8,8) and (4,8), as expected. For the CBH system with $U/J=1$, we find good but not perfect agreement, namely, $f_s = 0.9745$ for (8,8) y $f_s = 0.985$ for (4,8), which is consistent with the results of the previous subsection. The small departure from unity in the CBH case is probably due to the fact that we are not working in the thermodynamic limit. 

For KDBH, with $\kappa=0.6$, we obtain substantially lower values, namely, $f_s = 0.371$ for (8,8) and $f_s = 0.093$ for (4,8). It is important to note that the result in Ref. \cite{98LE} proving that $f_s$ must be unity at zero temperature does not apply to the KDBH model because the proof relies essentially of the existence of single-particle kinetic energy. 

We also note that the relatively small value of $f_s$ is consistent with the very different current-phase profile shown in Figs. \ref{fig:current_vs_phi}c and \ref{fig:current_condensate}b as compared with their CBH counterparts, which is reminiscent of weak superfluidity. The fact that in a homogeneous ring we obtain a current-flux relation that is qualitatively similar to that of a SQUID operating in the ideal Josephson regime [$\langle I \rangle \sim \sin (L\phi)$], is another remarkable feature of the KDBH system.

The high sensitivity of $f_s$ to the density (a factor of 4 increase as the density is doubled) is consistent with the results discussed in the previous subsection.

An interesting point is that a one-dimensional system of free fermions may naively yield a nonzero superfluid fraction; specifically, $f_s=1$ in the low density limit and $f_s=2/\pi$ at half-filling. By contrast, the calculation of the superconducting kernel (which is proportional to the superfluid fraction) appearing in the London equation yields $f_s=0$ for free fermions in any dimension \cite{96TI}. The reason for the discrepancy is that those nonzero results ignore the equilibrium distribution of excitations (pure electron-hole pairs) in the presence of a flux or vector potential. Such a ``paramagnetic'' contribution from the quasiparticles exactly cancels the nonzero ``diamagnetic'' contribution \cite{96TI}. Thus we can state that independent bosons show the HF effect but independent fermions do not.

\section{\label{sec:persistent} Metastability of supercurrents}

Metastable currents in a ring with a barrier and not threaded by a flux are one of the features traditionally associated with superfluidity. In the absence of flux, there are non-stationary but long-lived configurations that carry a nonzero steady current. In the presence of some imperfection in the ring (which is usually the case), such current-carrying states can decay due to current reversal or to the spontaneous excitation of quasiparticles when the flow velocity is high enough.

Current reversal due to a single impurity is negligible in the thermodynamic limit. The resulting metastability is further enhanced by the presence of repulsive interactions, which translates into an attractive interaction in momentum space that favors the macroscopic occupation of a single momentum state 
\cite{12HE, 19PI}. 

With our small ring model we can explicitly analyze the mechanism of current reversal, which is particularly clear in the case of the conventional BH system. For small but nonzero flux we identify current-carrying excited states of the mean-field type, which are characterized by the macroscopic occupation of a opposite nonzero momenta, for instance, $\pm \pi/4$. For zero flux and zero barrier strength, these states are degenerate and genuinely stationary \footnote{The case $\varepsilon=0$ with $\phi = 0$ must be handled with care, since exact diagonalization may yield arbitrary states within the degenerate Kramers doublet. To avoid this problem, we work with a tiny flux that breaks the degeneracy without appreciably shifting the energy levels.}. For $\phi=0$ and $0< \varepsilon \ll 1$, the weak barrier breaks the degeneracy by mixing the two states and yielding a gap that can be identified with the inverse lifetime against current reversal of the metastable current-carrying states.

\begin{figure}[h!]
	\centering
	\includegraphics[width=1\linewidth]{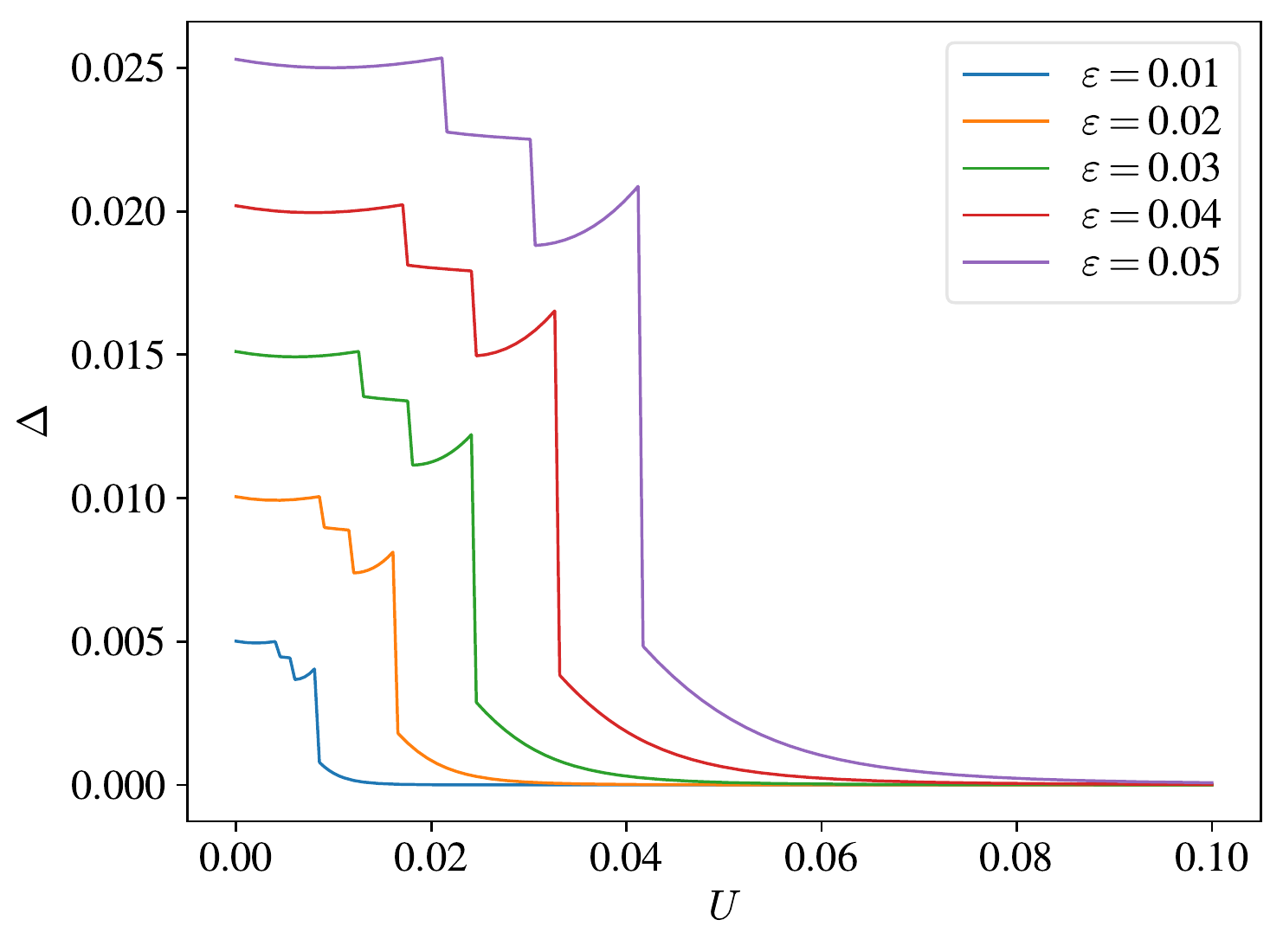}
	\caption{Gap energies $\Delta$ between the states of the CBH Hamiltonian \eqref{eq:Hcbh} of the form  $\ket{\Psi_{\pm}}\simeq(\ket{N_{\pi/4}}\pm\ket{N_{-\pi/4}})/\sqrt{2}$ plotted as a function of the interaction energy $U$ for several values of the impurity strength $\varepsilon$. Here, $N=L=8$, $J=1$, $\phi=0$. The discontinuities are ascribed to spurious finite-size effects.}
	\label{fig:gap_U}       
\end{figure}

Since repulsive interactions in real space favor the occupation of a single momentum state \cite{12HE}, we can expect the lifetime to grow (or the gap to decrease) with increasing $U>0$. This is what we actually find for the CBH model, as shown in Fig. \ref{fig:gap_U} for different impurity strengths. The clear global trend is that the gap $\Delta$ (proportional to the inverse lifetime) decreases with increasing $U$ and decreasing $\varepsilon$. The discontinuities in $\Delta$ as a function of $U$ arise from spurious finite-size effects.

A similar study in the KDBH system is more cumbersome to perform because there are many states between the symmetric and antisymmetric combinations of the metastable, mean-field--like current-carrying states. Moreover, the global role played by $U$ [see Eqs. \eqref{eq:Hkdbh}-\eqref{eq:Hkdbh2}] prevents its use as an interesting tuning parameter.

The other mechanism of metastable current decay is the spontaneous generation of quasiparticles when the flow velocity exceeds the Landau critical velocity which, in the presence of an acoustic mode, corresponds to the speed of sound. This mechanism remains relevant, and even more efficient, in the thermodynamic-limit, because of the presence of a continuum of available momenta and energies in the spectrum of low-lying excited states.

In Ref. \cite{18PI} some Luttinger-liquid theorems were invoked to indirectly prove the existence of superfluidity, which is ultimately associated with the existence of an acoustic mode. Here we choose to explore the excitation spectrum directly by calculating the dynamic form factor \cite{96PI}
\begin{equation}
	\label{eq:B}
	S(q,\omega)=\sum_{n=1}^{\text{dim}(\mathscr{H})-1}\left|
	\braket{\Psi_n|\rho^\dagger_q|\Psi_0}\right|^2\delta(\omega - \omega_{n0}) \, ,
\end{equation}
where $\rho_q$ is given in Eq. \eqref{rho-q} and $\omega_{n0} = E_n - E_0$, with $E_0$ the energy of the ground state $\ket{\Psi_0}$ and $E_n$ the energy of $n$-th excited state $\ket{\Psi_n}$.

\begin{figure}
\centering
\includegraphics[width=1\linewidth]{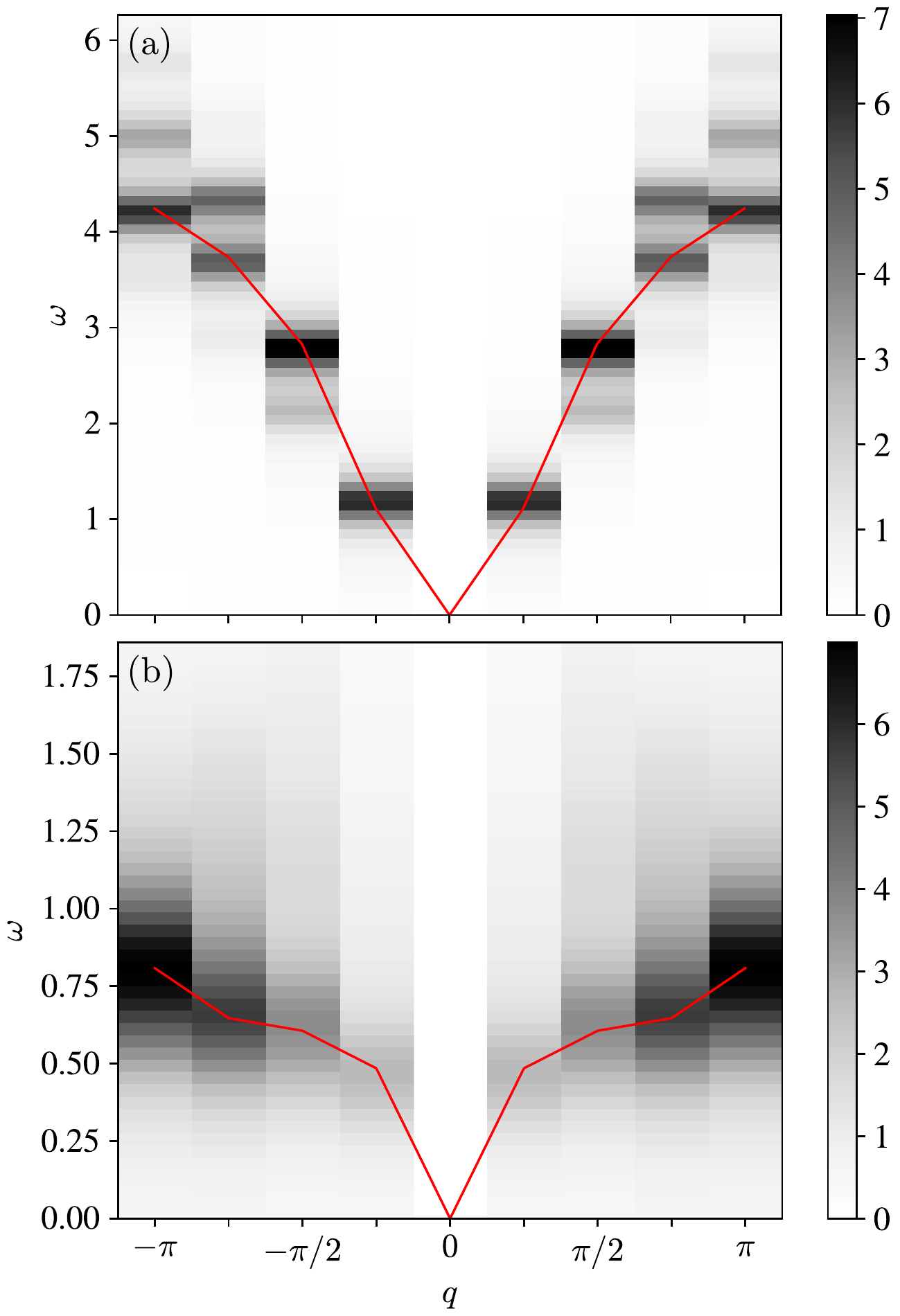}
\caption{Dynamic form factor \eqref{eq:B} for (a) the CBH Hamiltonian \eqref{eq:Hcbh} and (b) the KDBH Hamiltonian \eqref{eq:Hkdbh}, both in the superfluid regimes ($J/U=1$ and $\kappa=0.6$, respectively) and for  $N=L=8$. The flux is $\phi=0$ because a tiny nonzero flux is not needed due to completeness within the subspace of degeneracy.} 
\label{fig:S} 
\end{figure}

The numerical results for the number-conserving excitation spectrum are shown in Fig. \ref{fig:S} for the CBH and the KDBH models in their respective superfluid regimes. The frequency dependence is smoothed out with a convolution procedure similar to that employed in Ref. \cite{18PI}. The existence of an acoustic mode is clear in both cases. Figure \ref{fig:S}b provides direct numerical evidence of the existence of superfluidity in the KDBH model. This superfluidity is qualitatively different from that of the CBH model in several respects which we have discussed. We identify the existence of excited states with a dispersion relation similar to that of conventional quasiparticles but which actually are excitations shared by the two branches of the cat-type ground state.

For CBH (see Fig. \ref{fig:S}a), it is interesting to note that, for large momenta, the spectral peaks split into two. This is a system whose elementary excitations are well explained by Bogoliubov theory. In a tight-binding lattice, the acoustic relation does not hold for momenta near the boundary of the Brillouin zone. As a consequence, a given momentum of the elementary excitation can be due to configurations with different total energy, which results in a double peak in the dispersion relation.

It is interesting to analyze the dependence of the sound velocity $c$ with respect to the total particle density $n=N/L$. For CBH we expect $c \sim \sqrt{n}$, if we take the Bogoliubov approximation in the thermodynamic limit as a reference.  We lack a similar {\it a priori} expectation for the KDBH system. For CBH, Fig. \ref{fig:S}a reveals $c=1.41$ for $(N,L)=(8,8)$, while the same calculation for the case $(N,L)=(4,8)$ produces the result $c=1.16$ (not shown). This yields a ratio $1.22$, not very different from the reference value $\sqrt{2}$. A similar calculation for the KDBH model yields the ratio $0.61/0.25=2.44$. The fact that speed of sound is considerably more sensitive to the density than in the CBH case, is consistent with the results reported in section \ref{subsec:n-dependence} for the dependence of the total current on the boson density (see Fig. \ref{fig:ratio_current}).

Although a standard, Bogoliubov-type theory of the elementary excitation does not apply to the KDBH system \cite{19PI}, some physics insights are still possible. When a pair of total momentum $\pi$ is ``created'' (borrowing from the Bogoliubov mean-field language), it is not possible to distinguish from which cat branch the two atoms come, as both branches provide the same total momentum, namely, $\pi$, which is equivalent to $-\pi$. This fundamental inability to distinguish between the two cat branches lies at the root of the unusually robust cat features of the ground state.

\begin{figure}
\centering
\includegraphics[width=1\linewidth]{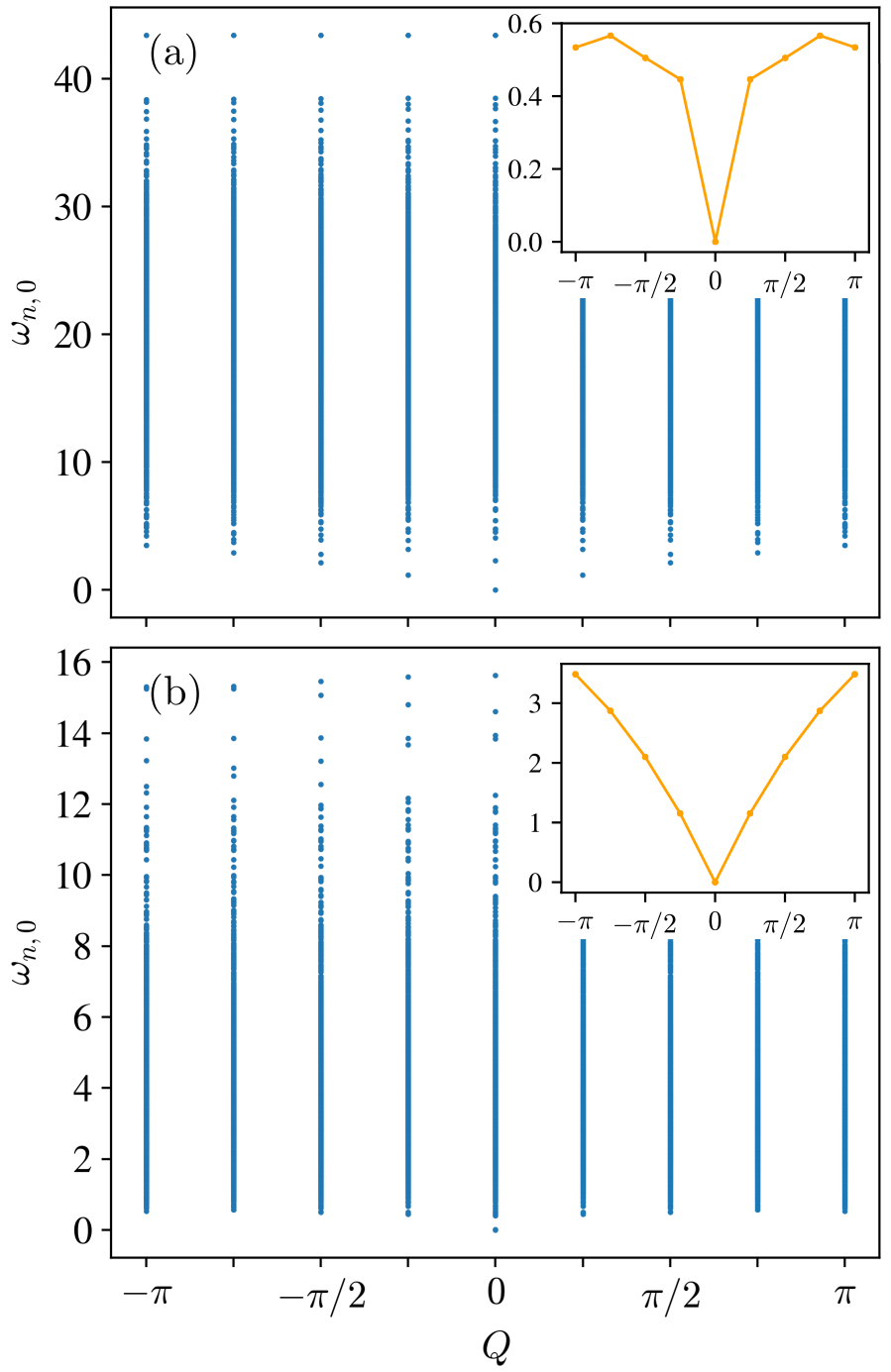}
\caption{Excitation energies $\omega_{n0}$ for each value of the total momentum $Q$ \eqref{eq:Q}, for (a) the CBH model \eqref{eq:Hcbh} and (b) the KDBH system \eqref{eq:Hkdbh} in (b). Parameters as in Fig. \ref{fig:S}. The insets show the lowest values of $\omega_{n0}$ (i.e., $\omega_{10}$) in each $Q$ sector. To have $Q$ taking only the allowed momentum values, it is important to introduce a tiny flux ($\phi=10^{-6}$) to slightly break the degeneracies and thus avoid combinations of degenerate states that are not eigenstates of the total momentum.}
\label{fig:omega_Q} 
\end{figure}

For completeness and comparison, it is interesting to plot the entire energy spectrum as a function of the total momentum
\begin{align}
	\label{eq:Q}
	Q = \sum_{\ell=0}^{L-1}\left(\frac{2\pi\ell}{L}\right)n_\ell\ \quad(\text{mod }2\pi).
\end{align}
The results are shown in Fig. \ref{fig:omega_Q} for both CBH and KDBH. In both cases, we provide an inset that tracks the ground-state energy of the various $Q$ sectors. It is interesting to compare those curves with the dynamic form factor, which for a given $q$ involves many excited states. For this reason, $S(q,\omega)$ peaks at energies higher than the ground-state energy of the same momentum sector.

\section{\label{sec:conclusions}Conclusions}

We have investigated the nature of the superfluidity in a kinetically driven one-dimensional boson system described, in the absence of driving, by the Bose-Hubbard model. Kinetic driving is a form of Floquet engineering where the hopping energy is made to oscillate with zero time-average. The resulting effective dynamics is that of a system where first-order single-particle hopping processes are suppressed and only higher-order processes involving more than one particle are allowed. Thus, kinetic driving is an alternative route to the design of flat-band systems. 

In our study, we compare the dynamics of this highly correlated system with that of the conventional (undriven) Bose-Hubbard model. In both cases, we employ exact diagonalization, which limits the system's size and particle number but permits detailed studies. The main message is that superfluidity is possible in a flat-band system and that, in many regards, it is qualitatively different from the conventional superfluidity based on autonomous single-particle hopping.

We have paid attention to the two different facets or definitions of superfluidity, namely, the Hess-Fairbank effect and the metastability of supercurrents, with explicit inclusion of an impurity in some cases. For the HF effect, we have studied the dependence of the kinetically driven Bose-Hubbard (KDBH) ground-state current on an effective flux. When comparing with the conventional Bose-Hubbard (CBH) system, we find a number of important differences.

The superfluid current of the two branches of the cat-like ground state is identical despite their crystal momenta being opposite. Moreover, the interference between the various collective components of the ground-state wave function is essential to produce the ``diamagnetic'' current. The diagonal contribution of the ideal cat part to the current  is zero. The analogous contributions of the other two components (shared condensate and reduction cloud) are nonzero but considerably smaller than those resulting from their interference with the ideal cat. This contrasts sharply with the CBH case, where the single condensate carries diagonally most of the flux-dependent current and its crossed contribution with the depletion cloud is zero.

Another important feature is that, in the presence of zero or weak impurities, the KDBH shows a roughly sinusoidal dependence of the current on the flux, instead of the characteristic linear behavior of the CBH system. This means that, even in the absence of impurities, the KDBH behaves like CBH does only when a weak link (strong impurity) is present in the ring.

The dependence of the diamagnetic current on the density is also quite anomalous. While CBH shows the expected linear dependence, the KDBH model displays a higher sensitivity. Similar results are obtained for the superfluid fraction, which is close to unity in the CBH case and appreciably smaller for the KDBH system.

Defining metastable supercurrents as those that are long lived in the presence of an impurity and without the help of a flux, we have studied the current reversal due to the presence of an impurity, a mechanism that can be important in small rings. For the CBH case we have explicitly shown that the inverse lifetime against current reversal increases with the impurity strength and decreases with the real-space repulsive interaction.

To investigate the robustness of the supercurrent against the spontaneous creation of excitations, we have computed the dynamic form factor and found that the KDBH shows an acoustic mode with a speed of sound that, like for CBH, is to be identified with the Landau critical velocity. This ensures the existence of metastable supercurrents in the thermodynamic limit.

One may wonder whether our conclusion on the viability of superfluidity in the KDBH model is applicable to other flat-band systems. This question cannot be addressed in general, but it seems reasonable to expect that, as long as bosons can hop in pairs between nearest neighbors, they have the potential to form a superflowing condensate. However, if that is the only elementary process in real space, such a moving condensate will typically have a momentum per particle close to zero and thus will be not so different from a conventional superfluid, at least in this regard. In any case, we note that a full theoretical claim on the superfluidity of a flat-band system must be substantiated by an explicit check of its resilience against impurities in the presence of a flux, and by arguments supporting the existence of a nonzero Landau critical velocity.

In summary, we have found that a kinetically-driven boson system exhibits a qualitatively new type of superfluidity which is entirely based on the correlated motion of strongly interacting particles. Within the framework of a tractable model, our work proves that superfluidity in a flat band system is feasible. Floquet engineering based on the driving of the hopping energy opens new avenues in quantum many-body physics that may include higher-dimensional boson and fermion systems with possibly surprising properties.

\acknowledgments
We would like to thank Miguel A. Cazalilla, Anthony J. Leggett, 
Juan R. Mu\~noz de Nova, and Sandro Stringari, for valuable discussions. This work has been supported by Spain's Agencia Estatal de Investigaci\'on through Grant No. FIS2017-84368-P and by Universidad Complutense de Madrid through Grant No. FEI-EU-19-12. One of us (FS) would like to acknowledge the hospitality of the Sloane Physics Laboratory at Yale University, where part of this work was done.

\bibliography{sf_paper}

\end{document}